\documentclass[reprint,aip,jcp]{revtex4-1}
\draft

\usepackage{amsmath}
\usepackage{amsfonts}
\usepackage{amssymb}
\usepackage{amsxtra}
\usepackage{floatflt}
\usepackage{comment}
\usepackage[pdftex]{graphicx}
\usepackage{enumerate}
\usepackage{float}
\usepackage[normalem]{ulem}
\usepackage{color}

\usepackage{mathtools}
\graphicspath{{./figs/}}
\usepackage{marvosym}
\usepackage{times}
\usepackage{bm}
\usepackage{colortbl} 
\usepackage{multirow}
\usepackage{graphicx}
\usepackage{nameref}
\usepackage[usenames,dvipsnames,svgnames,table]{xcolor}
\usepackage{float}
\usepackage[usestackEOL]{stackengine}

\usepackage{tabularx}

\newcommand{\ly}{\left\{}
\newcommand{\ry}{\right\}}

\newcommand{\lp}{\left(}
\newcommand{\rp}{\right)}

\newcommand{\eg}{e.g.~}
\newcommand{\ie}{i.e.~}

\newcommand{\figref}[1]{Figure \ref{#1}}

\newcommand{\ex}{E_{1}}
\newcommand{\cx}{C_{1}}
\newcommand{\px}{P_{1}}
\newcommand{\rx}{\text{mRNA}_{1}}
\newcommand{\ez}{E_{2}}
\newcommand{\cz}{C_{2}}
\newcommand{\pz}{P_{2}}
\newcommand{\rz}{\text{mRNA}_{2}}
\newcommand{\seq}{E^{*}}

\newcommand{\exb}{E_1}
\newcommand{\cxb}{C_1}

\newcommand{\ezb}{E_2}
\newcommand{\czb}{C_2}

\newcommand{\seqb}{E^{*}}

\newcommand{\etotx}{E_{\text{t},1}}
\newcommand{\etotz}{E_{\text{t},2}}

\newcommand{\evec}{E}
\newcommand{\pvec}{P}
\newcommand{\pmean}{\bar{P}}

\newcommand{\kfx}{k_{\text{f},1}}
\newcommand{\kfj}{k_{\text{f},j}}
\newcommand{\kfxx}{k_{\text{f},1}}
\newcommand{\kbx}{k_{\text{b},1}}
\newcommand{\kbj}{k_{\text{b},j}}
\newcommand{\kcx}{k_{\text{cat},1}}
\newcommand{\krx}{k_{\text{rev},1}}
\newcommand{\kconsx}{k_{\text{c},1}}

\newcommand{\kfz}{k_{\text{f},2}}
\newcommand{\kbz}{k_{\text{b},2}}
\newcommand{\kcz}{k_{\text{cat},2}}
\newcommand{\krz}{k_{\text{rev},2}}
\newcommand{\kconsz}{k_{\text{c},2}}

\newcommand{\kseq}{k_{\text{sq}}}
\newcommand{\kunseq}{k_{\text{rsq}}}

\newcommand{\ktxx}{k_{\text{tx},1}}
\newcommand{\ktlx}{k_{\text{tl},1}}
\newcommand{\kdegrx}{k_{\text{deg},1}}
\newcommand{\ktxz}{k_{\text{tx},2}}
\newcommand{\ktlz}{k_{\text{tl},2}}
\newcommand{\kdegrz}{k_{\text{deg},2}}

\newcommand{\kpa}{\kappa_{\text{S}}}
\newcommand{\kpb}{\kappa_{1}}
\newcommand{\kpc}{\kappa_{2}}
\newcommand{\kpd}{\kappa_{3}}

\newcommand{\kma}{K_{\text{m},S}}
\newcommand{\kmb}{K_{\text{m},1}}
\newcommand{\kmc}{K_{\text{m},2}}
\newcommand{\kmd}{K_{\text{m},3}}

\newcommand{\enzyme}{E}
\newcommand{\complex}{C}
\newcommand{\product}{P}
\newcommand{\substrate}{S}
\newcommand{\etotal}{E_{\text{total}}}

\newcommand{\enzymeb}{E}
\newcommand{\complexb}{C}

\newcommand{\kfor}{k_\text{f}}
\newcommand{\kback}{k_\text{b}}
\newcommand{\kcat}{k_\text{cat}}
\newcommand{\krev}{k_\text{rev}}
\newcommand{\kdil}{\delta}
\newcommand{\kexp}{k_\text{c}}
\newcommand{\ktransc}{k_{\text{tx}}}
\newcommand{\ktransl}{k_{\text{tl}}}
\newcommand{\kdegr}{k_{\text{deg}}}

\begin{document}
	
\title{Computation of single-cell metabolite distributions using mixture models}
\author{Mona K. Tonn}

\affiliation{Department of Mathematics, Imperial College London, SW7 2AZ, London, United Kingdom}

\author{Philipp Thomas}
\affiliation{Department of Mathematics, Imperial College London, SW7 2AZ, London, United Kingdom}

\author{Mauricio Barahona}
\affiliation{Department of Mathematics, Imperial College London, SW7 2AZ, London, United Kingdom}

\author{Diego A. Oyarz\'un}
\email[]{Corresponding author: d.oyarzun@ed.ac.uk}
\email[]{This work was partly funded by the Human Frontier Science Program through a Young Investigator Grant (RGY0076-2015) awarded to D.O., a UKRI Future Leaders Fellowship (MR/T018429/1) awarded to P.T., and the EPSRC Centre for Mathematics of Precision Healthcare (EP/N014529/1) awarded to M.B.}
\affiliation{School of Biological Sciences, University of Edinburgh, United Kingdom}
\affiliation{School of Informatics, University of Edinburgh, United Kingdom}
%\date{\today}

\begin{abstract}
Metabolic heterogeneity is widely recognised as the next challenge in our understanding of non-genetic variation. A growing body of evidence suggests that metabolic heterogeneity may result from the inherent stochasticity of intracellular events. However, metabolism has been traditionally viewed as a purely deterministic process, on the basis that highly abundant metabolites tend to filter out stochastic phenomena. Here we bridge this gap with a general method for prediction of metabolite distributions across single cells. By exploiting the separation of time scales between enzyme expression and enzyme kinetics, our method produces estimates for metabolite distributions without the lengthy stochastic simulations that would be typically required for large metabolic models. The metabolite distributions take the form of Gaussian mixture models that are directly computable from single-cell expression data and standard deterministic models for metabolic pathways. The proposed mixture models provide a systematic method to predict the impact of biochemical parameters on metabolite distributions. Our method lays the groundwork for identifying the molecular processes that shape metabolic heterogeneity and its functional implications in disease.% such as antibiotic tolerance and tumour heterogeneity in cancer.
\end{abstract}

\pacs{}

\maketitle

\section{Introduction}

Non-genetic heterogeneity is a hallmark of cell physiology. Isogenic cells can display markedly different phenotypes as a result of the stochasticity of intracellular processes and fluctuations in environmental conditions. Gene expression variability, in particular, has received substantial attention thanks to robust experimental techniques for measuring transcripts and proteins at a single-cell resolution \cite{Golding2005,Taniguchi2010}. This progress has gone hand-in-hand with a large body of theoretical work on stochastic models to identify the molecular processes that affect expression heterogeneity\cite{Swain2002,Raj2008,Thomas2014,Tonn2019,Dattani2017}. 

In contrast to gene expression, our understanding of stochastic phenomena in metabolism is still in its infancy. Traditionally, cellular metabolism has been regarded as a deterministic process on the basis that metabolites appear in large numbers that filter out stochastic phenomena\cite{Heinemann2011}. But this view is changing rapidly thanks to a growing number of single-cell measurements of metabolites and co-factors\cite{Esaki2015, Bennett2009, Xiao2016, Mannan2017, Lemke2011, Yaginuma2014, Imamura2009, Paige2012, Ibanez2013} that suggest that cell-to-cell metabolite variation is much more pervasive than previously thought. The functional implications of this heterogeneity are largely unknown but likely to be substantial given the roles of metabolism in many cellular processes, including growth\cite{Weisse2015}, gene regulation\cite{Lempp2019}, epigenetic control\cite{Loftus2016} and immunity\cite{Reid2017}. For example, metabolic heterogeneity has been linked to bacterial persistence\cite{Shan2017,Radzikowski2017}, a dormant phenotype characterised by a low metabolic activity, as well as antibiotic resistance \cite{Deris2013} and other functional effects \cite{Vilhena2018}. In biotechnology applications, metabolic heterogeneity is widely recognised as a limiting factor on metabolite production with genetically engineered microbes \cite{Schmitz2017,Binder2017,Liu2018}.

A key challenge for quantifying metabolic variability is the difficulty in measuring cellular metabolites at a single-cell resolution\cite{Amantonico2010,Takhaveev2018, Wehrens2018}. As a result, most studies use other phenotypes as a proxy for metabolic variation, \eg enzyme expression levels\cite{Kotte2014,vanHeerden2014}, metabolic fluxes\cite{Schreiber2016} or growth rate\cite{Kiviet2014,Simsek2018}. From a computational viewpoint, the key challenge is that metabolic processes operate on two timescales: a slow timescale for expression of metabolic enzymes, and a fast timescale for enzyme catalysis. Such multiscale structure results in stiff models that are infeasible to solve with standard algorithms for stochastic simulation\cite{Gillespie2007}. Other strategies to accelerate stochastic simulations, such as $\tau$-leaping, also fail to produce accurate simulation results due to the disparity in molecule numbers between enzymes and metabolites\cite{Tonn2020}. These challenges have motivated a number of methods to optimise stochastic simulations of metabolism\cite{Puchaka2004,Cao2005Accelerated,Labhsetwar2013,Lugagne2013,Murabito2014}. Most of these methods exploit the timescale separation to accelerate simulations at the expense of some approximation error. This progress has been accompanied by a number of theoretical results on the links between molecular processes and the shape of metabolite distributions\cite{Levine2007,Gupta2017, Oyarzun2015,Tonn2019}. Yet to date there are no general methods for computing metabolite distributions that can handle inherent features of metabolic pathways such as feedback regulation, complex stoichiometries, and the high number of molecular species involved. 

In this paper we present a widely applicable method for approximating single-cell metabolite distributions. Our method is founded on the timescale separation between enzyme expression and enzyme catalysis, which we employ to approximate the stationary solution of the chemical master equation. The approximate solution takes the form of mixture distributions with: (i) mixture weights that can be computed from models for gene expression or single-cell expression data, and (ii) mixture components that are directly computable from deterministic pathway models. The resulting mixture model can be employed to explore the impact of biochemical parameters on metabolite variability. We illustrate the power of the method in two exemplar systems that are core building blocks of large metabolic networks. Our theory provides a quantitative basis to draw testable hypotheses on the sources of metabolite heterogeneity, which together with the ongoing efforts in single-cell metabolite measurements, will help to re-evaluate the role of metabolism as an active source of phenotypic variation.

%\vfill

%%%%%%%%%%%%%%%%%%%%%%%%%%%
%%%%%%%%%%%%%%%%%%%%%%%%%%%
%%%%%%%%%%%%%%%%%%%%%%%%%%%

\section{General method for computing metabolite distributions}
We consider metabolic pathways composed of enzymatic reactions interconnected by sharing of metabolites as substrates or products. In general, we consider models with $M$ metabolites $P_{i}$ with $i \in \left\{1,2,...,M \right\}$ and $N$ catalytic enzymes $E_{j}$ with $j \in \left\{1,2,...,N \right\}$. A typical enzymatic reaction has the form
\begin{align}\label{eq:genericreaction}
    P_i + E_j \xrightleftharpoons[\kbj]{\kfj} C_j \xrightleftharpoons[k_{\text{rev},j}]{k_{\text{cat},j}} P_{k} + E_j,
\end{align}
where $P_i$ and $P_k$ are metabolites, and $E_j$ and $C_j$ are the free and substrate-bound forms of the enzyme. The parameters $(\kfj,\kbj)$ and $(k_{\text{cat},j},k_{\text{rev},j})$ are positive rate constants specific to the enzyme. In contrast to traditional metabolic models, where the number of enzyme molecules is assumed constant, here we explicitly model enzyme expression and enzyme catalysis as stochastic processes. %To this end, we assume that a separate stochastic process controls the number of enzymes available to the reaction. 
Our models also account for dilution of molecular species by cell growth and consumption of the metabolite products by downstream processes.

Though in principle one can readily write a Chemical Master Equation (CME) for the marginal distribution $\mathbf{P}(P_{1},P_{2},...P_{M})$ given the pathway stoichiometry, analytical solutions of the CME are tractable only in few special cases. To overcome this challenge, we propose a method for approximating metabolite distributions that can be applied in a wide range of metabolic models. We first note that using the Law of Total Probability, the marginal distribution $\mathbf{P}(P_{1},P_{2},...P_{M})$ can be generally written as:
% \begin{align}\label{eq:law_total_prob_general}
% \mathbf{P}(\pvec) = \sum_{\evec} \underbrace{\mathbf{P}(\evec)}_{\substack{\text{gene}\\ \text{expression}}} \times  \underbrace{\mathbf{P}(\pvec|\evec)}_\text{catalysis},
% \end{align}
\begin{align}\label{eq:law_total_prob_general}
\mathbf{P}(\pvec) = \sum_{\evec} \mathbf{P}(\evec) \times  \mathbf{P}(\pvec|\evec),
\end{align}
where $\pvec=(P_{1},P_{2},...P_{M})$ and  $\evec = \left(E_{1}, E_{2},..., E_{N} \right)$ are the vectors of metabolite and enzyme abundances, respectively. The equation in \eqref{eq:law_total_prob_general} describes the metabolite distribution in terms of fluctuations in gene expression, comprised in the distribution $\mathbf{P}(\evec)$, and fluctuations in reaction catalysis, described by conditional distribution $\mathbf{P}(\pvec|\evec)$. 

A key observation is that Eq.~\eqref{eq:law_total_prob_general} corresponds to a mixture model with weights $\mathbf{P}(\evec)$ and mixture components $\mathbf{P}(\pvec|\evec)$. To compute the mixture weights and components, we make use of the timescale separation between gene expression and metabolism. Gene expression operates on a much slower timescale than catalysis\cite{Cao2005Accelerated,Levine2007,Kuntz2013}, with protein half-lives typically comparable to cell doubling times and catalysis operating in the millisecond to second range. Therefore, in the fast timescale of catalysis we can write a conservation law for the total amount of each enzyme (free and bound):
\begin{align}\label{eq:consj}
    E_{\text{t},j} &= E_j + C_j,
\end{align}
where $E_{\text{t},j}$ is the total number of enzymes $E_j$. Note that since our models integrate enzyme kinetics with enzyme expression, the variables $E_{\text{t},j}$ follow their own, independent stochastic dynamics. It is important to note that in our approach, the conservation relation in \eqref{eq:consj} holds only in the fast timescale of catalysis. This contrasts with classic deterministic models for metabolic reactions, which typically focus on the fast catalytic timescale and assume enzymes as constant model parameters\cite{cornish-bowden04a}. 

\begin{figure}[H]
\begin{center}
    \includegraphics{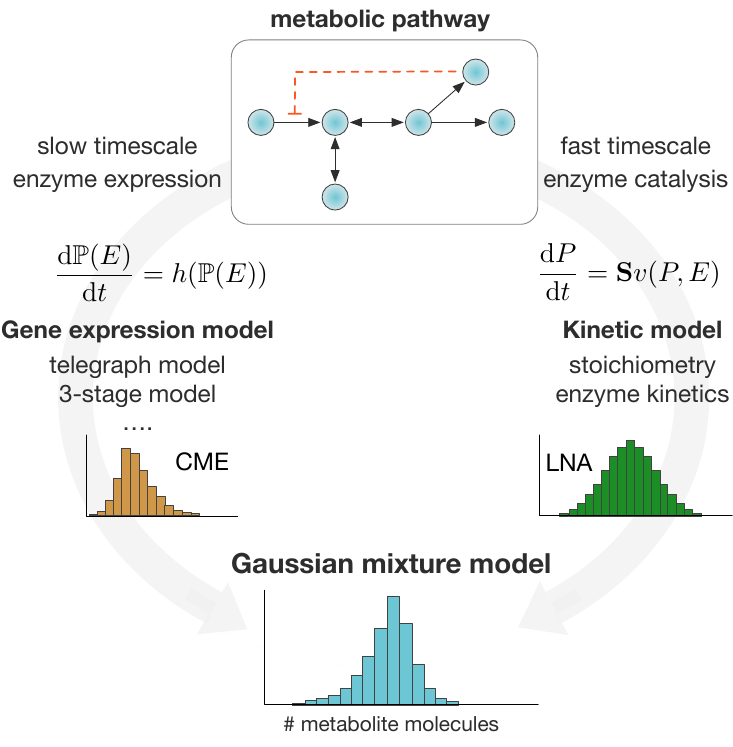}
\end{center}
\caption{\textbf{Computation of single-cell metabolite distributions with Gaussian mixture models.} We exploit the separation of timescales to compute the weights and components of the mixture model in Eq.~\eqref{eq:law_total_prob_general}. Mixture weights are computed as stationary solutions to the Chemical Master Equation (CME) for a chosen model for stochastic enzyme expression. The mixture components are computed via the Linear Noise Approximation\cite{Elf2003} (LNA) applied to the pathway ODE model. This provides Gaussian approximations to the stationary metabolite distribution conditioned on the enzyme state. Overall, the method produces a Gaussian mixture model for metabolite distributions that can be applied in a wide range of metabolic pathways.} 
\label{fig:concept}
\end{figure}

As a result of the separation of timescales, the weights and components of the mixture in Eq.~\eqref{eq:law_total_prob_general} can be computed separately. The mixture weights $\mathbf{P}(\evec)$, in particular, can be computed as solutions of a stochastic model for enzyme expression\cite{Raj2008}, or taken from single-cell measurements of enzyme expression. The mixture components $\mathbf{P}(\pvec|\evec)$, on the other hand, can be estimated with the Linear Noise Approximation\cite{vanKampen1992,Elf2003} (LNA) on the basis that metabolites appear in large numbers. In \figref{fig:concept} we illustrate a schematic of the proposed method.

We thus propose the following procedure for computing single-cell metabolite distributions:
\begin{enumerate}
    \item Starting from the mixture model in Eq.~\eqref{eq:law_total_prob_general}, compute the enzyme distribution $\mathbf{P}(\evec)$ from a stochastic model for gene expression, either analytically (if possible) or numerically with Gillespie's algorithm.
    
    \item To approximate the mixture components $\mathbf{P}(\pvec|\evec)$ with the LNA, compute the steady state solution $\pmean$ of the deterministic rate equation for each enzyme state $\evec$:
    \begin{align}\label{eq:ODEgral}
        \mathbf{S}v(\pmean,\evec) =0,
    \end{align}
    where $\mathbf{S}$ is the stoichiometric matrix and $v(\cdot)$ is the vector of deterministic reaction rates; for ease of notation we have assumed a unit cell volume, and hence the deterministic rates are equal to the propensities of the stochastic model. Note that due to the timescale separation, Eq.~\eqref{eq:ODEgral} must be solved assuming constant enzymes $\evec$, and its solution depends on the enzyme abundance, \ie $\pmean=\pmean(\evec)$
    .
    
    \item For each enzyme state $\evec$, compute the solution to the Lyapunov equation\cite{Elf2003}:
    \begin{align}\label{eq:lyapunov-general}
       A \Sigma + \Sigma A^{T} + B B^{T} = 0,
    \end{align}
    where $A$ is the Jacobian of \eqref{eq:ODEgral} evaluated at the steady state and $BB^{T}=\mathbf{S}\text{diag}\ly v \ry \mathbf{S}^{T}$. Note that, as in \eqref{eq:ODEgral}, the solution of the Lyapunov equation depends on the enzyme state, \ie $\Sigma=\Sigma(\evec)$.
    
    \item Following the LNA, approximate the mixture components  $\mathbf{P}(\pvec|\evec)$ as a multivariate Gaussian distribution with mean $\pmean$ and covariance matrix $\Sigma$.
    
    \item Combine the weights  $\mathbf{P}(\evec)$ and Gaussian components $\mathbf{P}(\pvec|\evec)$ through the mixture model in \eqref{eq:law_total_prob_general}.
\end{enumerate}

In the next sections we illustrate the effectiveness of our method in two exemplar systems.

\section{Reversible Michaelis-Menten reaction}\label{seq:case1}
\begin{figure}
\includegraphics{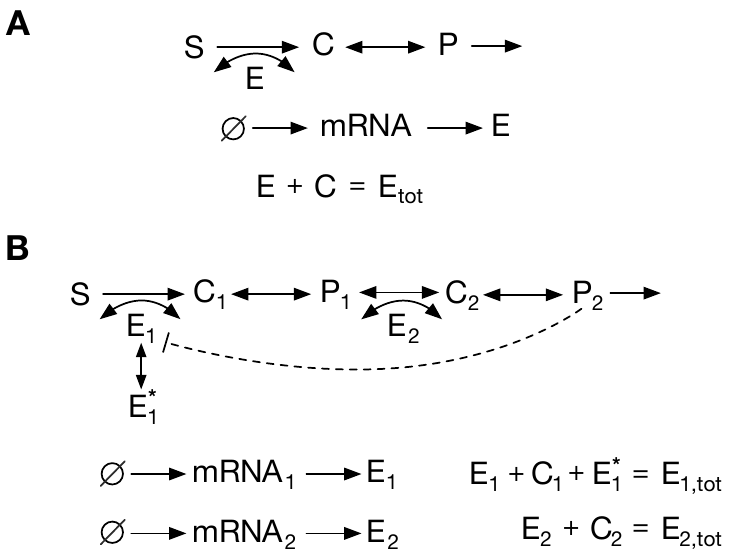}
\caption{\textbf{Exemplar metabolic systems}. (\textbf{A}) Reversible Michaelis-Menten reaction; the full set of reactions are shown in Eq.~\eqref{eq:revMM1}--\eqref{eq:twostage1}. The model accounts for reversible catalysis of a substrate $S$ into a product $P$. (\textbf{B}) Two-step pathway with noncompetitive end-product inhibition; the reactions are shown in Eq.~\eqref{eq:epi_PC_1}--\eqref{eq:epi_DIL}. The product ($P_2$) sequesters enzyme $E_1$ into an inactive form $E_1^*$, thereby reducing the rate of the first reaction. In both examples we assume a constant substrate $S$ and linear dilution of all chemical species. Enzymes are assumed to follow the two-stage model for gene expression\cite{Shahrezaei2008}, which includes species for the enzymatic mRNA and protein.}
\label{fig:casestudies}
\end{figure}

We first consider a stochastic model that integrates a reversible Michaelis-Menten reaction with a standard model for enzyme expression. As shown in \figref{fig:casestudies}A, the Michaelis-Menten mechanism includes reversible binding of four species: a metabolic substrate $\substrate$, a free enzyme $\enzyme$, a substrate-enzyme complex $\complex$ and a metabolic product $\product$. To model enzyme expression, we use the well-known two-stage scheme for transcription and translation\cite{Thattai2001,Shahrezaei2008} (\figref{fig:casestudies}A). The complete set of reactions is: 
\begin{align}
&\substrate + \enzyme \xrightleftharpoons[\kback]{\kfor} \complex \xrightleftharpoons[\krev]{\kcat} \product + \enzyme,\label{eq:revMM1}\\
&\emptyset  \xrightarrow[]{\ktransc} \text{mRNA} \xrightarrow[]{\ktransl} \text{mRNA}+ \enzyme,\label{eq:twostage1}\\
&\product \xrightarrow[]{\kexp} \emptyset ,\quad \text{mRNA} \xrightarrow[]{\kdegr}\emptyset,\label{eq:degrcons}\\
& \enzyme \xrightarrow[]{\kdil} \emptyset ,\quad \complex \xrightarrow[]{\kdil} \emptyset. \label{eq:revMM2}
\end{align}
The reactions in \eqref{eq:revMM1} correspond to a reversible Michaelis-Menten reaction as in \eqref{eq:genericreaction}, while reactions in \eqref{eq:twostage1} are the two-stage model for gene expression. We include four additional first-order reactions \eqref{eq:degrcons}--\eqref{eq:revMM2} to model consumption of the metabolite product with rate constant $\kexp$, mRNA degradation with rate constant $\kdegr$, and dilution of all model species with rate constant $\delta$. In what follows we assume that the substrate $\substrate$ remains strictly constant, for example to model cases in which the substrate represents an extracellular carbon source that evolves in much slower timescale than cell doubling times.

Since on the fast timescale of the catalytic reaction, the total number of enzymes can be assumed in quasi-stationary state\cite{cornish-bowden04a, Tonn2019}, we have that 
\begin{align}
    \etotal &= \enzyme+\complex, \label{eq:cons1}
\end{align}
and therefore the general mixture model in \eqref{eq:law_total_prob_general} can be written as:
\begin{align} \label{eq:mixmodel1}
\mathbf{P}(\product) = \sum_{\etotal = 0}^{\infty}\underbrace{\mathbf{P}(\etotal)}_{\substack{\text{enzyme}\\\text{distribution}}} \times \underbrace{\mathbf{P}(\product|\etotal)}_{\substack{\text{Gaussian}\\\text{from LNA}}}.
\end{align}

The mixture weights $\mathbf{P}(\etotal)$ can be computed from the stochastic model for gene expression in \eqref{eq:twostage1}. Under the standard assumption that mRNAs are degraded much faster than proteins\cite{Raj2008}, the stationary solution of the two-stage model can be approximated by a negative binomial distribution\cite{Shahrezaei2008}:
\begin{align}
    \mathbf{P}(\etotal) = \frac{\Gamma(a+\etotal)}{\Gamma(\etotal+1)\Gamma(a)}\lp\frac{b}{1+b}\rp^{\etotal} \frac{1}{(1+b)^a},\label{eq:negbin}
\end{align}
where $\Gamma$ is the Gamma function and the parameters are defined as the burst frequency $a = \ktransc\slash\kdil$ and burst size $b =\ktransl\slash\kdegr$.

To compute the mixture components $\mathbf{P}(\product|\etotal)$ with the LNA, we write the full system of deterministic rate equations (see \eqref{eq:ODE1} in Methods) for the three species $E$, $C$ and $P$. Note that in this case, we can further reduce the rate equations by (i) using the conservation law in \eqref{eq:cons1}, and (ii) assuming that the binding and unbinding reactions between $S$ and $E$ reach equilibrium faster than the product $P$, a condition that generally holds in metabolic reactions. After algebraic manipulations, the reduced ODE can be written as:
\begin{align}\label{eq:ODEred1}
    \frac{\text{d} \product }{\text{d} t} = f(\product,\etotal) - g(\product,\etotal) - \kexp \product
\end{align}
where
\begin{align}\label{eq:ODEred1_fun}
\begin{split}
    f(\product,\etotal) = \etotal \frac{\kcat \substrate\slash K_{\text{m}S}}{1 + \substrate/K_{\text{m}S} + \product/K_{\text{m}P}},\\
    g(\product,\etotal) = \etotal \frac{\kback \product\slash K_{\text{m}P}}{1 + \substrate/K_{\text{m}S} + \product/K_{\text{m}P}}
\end{split}
\end{align}
and the parameters are $K_{\text{m}S} = (\kback + \kcat)/\kfor$ and $K_{\text{m}P} = (\kback + \kcat)/\krev$.

The mean of each mixture component is simply given by the steady state solution of \eqref{eq:ODEred1}, which we denote as $\bar{\product}(\etotal)$. For a given enzyme abundance $\etotal$, the variance $\Sigma(\etotal)$ of each Gaussian component is given by the solution to the Lyapunov equation in  \eqref{eq:lyapunov-general}:
\begin{align}\label{eq:lyapred1}
    \Sigma(\etotal)
     &= \frac{1}{2}\frac{f(\bar{\product}(\etotal)) + g(\bar{\product}(\etotal)) + \kexp \bar{\product}(\etotal)}{ \kexp + g'(\bar{\product}(\etotal)) -f'(\bar{\product}(\etotal))},
\end{align}
where $f'$ and $g'$ are first-order derivatives. Combining the negative binomial in \eqref{eq:negbin} with the Gaussian components, we can rewrite Eq.~\eqref{eq:mixmodel1} to get a Gaussian mixture model for the metabolite: 
% \begin{align} \label{eq:mixmodel-case1}
% \mathbf{P}(\product) = \sum_{x = 0}^{\infty}\frac{b^a}{\Gamma(a)}x^{a-1}e^{-bx} \times \frac{1}{\sigma\sqrt{2\pi}}e^{-\frac{1}{2}\lp\frac{P-\pmean}{\sigma}\rp^2}.
% \end{align}
\begin{align} \label{eq:mixmodel-case1}
\mathbf{P}(\product) & = K\sum_{x = 0}^{\infty}\frac{1}{\Sigma(x)}\frac{\Gamma(a+x)}{\Gamma(x+1)}\lp\frac{b}{1+b}\rp^{x} e^{-\frac{1}{2}\lp\frac{P-\pmean(x)}{\Sigma(x)}\rp^2},
\end{align}
where both $\pmean(x)$ and $\Sigma(x)$ must be computed for each value of $x=\etotal$ in the summation. The normalization constant in \eqref{eq:mixmodel-case1} is
\begin{align}
 K = \frac{1}{\sqrt{2\pi}\Gamma(a)(1+b)^a}.
\end{align}

\begin{figure}[H]
\includegraphics{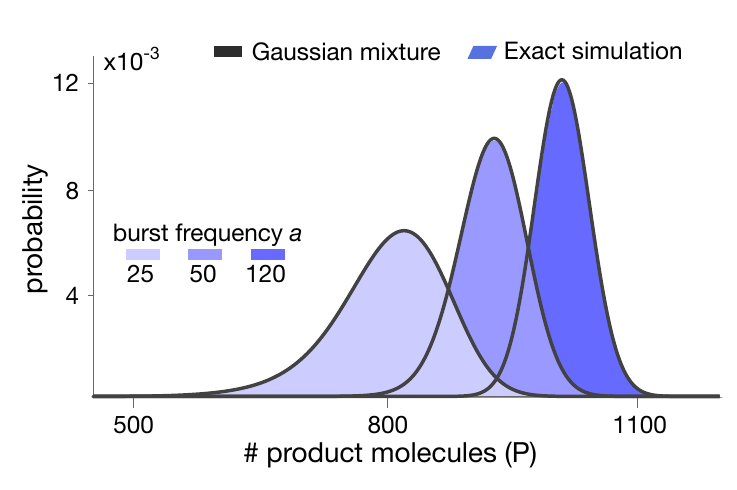}
\caption{\textbf{Stationary product distribution of a Michaelis-Menten reaction.} The proposed mixture model in \eqref{eq:mixmodel-case1} provides an excellent approximation for the metabolite distribution obtained with Gillespie's algorithm\cite{Gillespie2007}. Distributions were computed for varying values of the bursting parameter $a$, suggesting that high-frequency bursting tends to decrease metabolite skewness. All parameter values can be found in Table \ref{tab:casestudy1} in the Methods.}
\label{fig:casestudy1}
\end{figure}

In \figref{fig:casestudy1} we plot the mixture model \eqref{eq:mixmodel-case1} for realistic parameter values and compare this approximation with distributions computed from long runs of Gillespie simulations of the whole set of reactions \eqref{eq:revMM1}--\eqref{eq:revMM2}. The results indicate that the mixture model provides an excellent approximation of the metabolite distribution, even in the case of skewed or tailed distributions. In the next section we test our methodology in a more complex pathway with feedback regulation.

\section{Pathway with end-product inhibition}

A common regulatory motif in metabolism is end-product inhibition, in which a pathway enzyme can bind to its own substrate as well as the pathway product (see \figref{fig:casestudies}B). The product thus sequesters enzyme molecules, which reduces the number of free enzymes available for catalysis and slows done the reaction rate. To examine the accuracy of our method in this setting, we study a fully stochastic model for a two-step pathway with noncompetitive end-product inhibition:
\begin{align}
&\substrate + \ex \xrightleftharpoons[\kbx]{\kfx} \cx \xrightleftharpoons[\krx]{\kcx} \px + \ex \label{eq:epi_PC_1} \\
&\px + \ez \xrightleftharpoons[\kbz]{\kfz} \cz \xrightleftharpoons[\krz]{\kcz} \pz + \ez \label{eq:epi_PC_2} \\
&h \pz + \ex \xrightleftharpoons[\kunseq]{\kseq} \seq \label{eq:epi_SEQ} \\
&\emptyset \xrightarrow[]{\ktxx}\rx \xrightarrow[]{\ktlx} \rx + \ex \label{eq:epi_GE_1} \\
&\emptyset \xrightarrow[]{\ktxz}\rz \xrightarrow[]{\ktlz} \rz + \ez \label{eq:epi_GE_2} \\
&\px \xrightarrow[]{\kconsx} \emptyset,\quad
\pz \xrightarrow[]{\kconsz} \emptyset \label{eq:epi_PE_2} \\
& \rx \xrightarrow[]{\kdegrx} \emptyset,\quad \rz \xrightarrow[]{\kdegrz} \emptyset,\\
&\seq \xrightarrow[]{\kdil} \emptyset,\quad \ex \xrightarrow[]{\kdil} \emptyset ,\quad \ez \xrightarrow[]{\kdil} \emptyset ,\quad \cx \xrightarrow[]{\kdil} \emptyset ,\quad \cz \xrightarrow[]{\kdil} \emptyset \label{eq:epi_DIL} 
\end{align}
The two reactions in \eqref{eq:epi_PC_1} and \eqref{eq:epi_PC_2} are reversible Michaelis-Menten kinetics, sharing the intermediate metabolite $\px$ as a product and substrate, respectively. The end-product inhibition in \eqref{eq:epi_SEQ} consists of reversible binding between $h$ molecules of $\pz$ and the first enzyme $\ex$ into a catalytically-inactive complex $\seq$. The remaining model reactions in \eqref{eq:epi_GE_1}--\eqref{eq:epi_DIL} are analogous to the previous example in Section \ref{seq:case1}: reactions in \eqref{eq:epi_GE_1}--\eqref{eq:epi_GE_2} describe the two-stage model for expression of both enzymes, and with reactions \eqref{eq:epi_PE_2}--\eqref{eq:epi_DIL} we model first-order mRNA degradation, product consumption, and dilution by cell growth. For simplicity we also assume that both enzymes are independently expressed, but in general our method can also account for cases in which enzymes are co-expressed or co-regulated\cite{Chubukov2014}. The resulting model has two distinct pools of enzymes, which remain constant over the timescale of catalysis:
\begin{align}
\begin{split}
\etotx &= \ex + \seq + \cx, \\
\etotz &= \ez + \cz,
\end{split}\label{eq:cons2}
\end{align}
and therefore the mixture model in \eqref{eq:law_total_prob_general} becomes
\begin{align}\label{eq:GMM2}
\mathbf{P} (\px,\pz) &= \sum_{\etotx,\etotz} \underbrace{\mathbf{P}(\etotx,\etotz)}_{\substack{\text{enzyme}\\\text{distribution}}}\underbrace{\mathbf{P}(\px,\pz | \etotx,\etotz)}_{\substack{\text{Gaussian}\\\text{from LNA}}},
\end{align}
where the summation goes through all $(\etotx,\etotz)$ pairs. Since both enzymes are expressed independently, the enzyme distribution is the product of two negative binomials $\mathbf{P}(\etotx,\etotz)=\mathbf{P}(\etotx) \times \mathbf{P}(\etotz)$, each one analogous to the distribution in \eqref{eq:negbin}.
% \begin{align}\label{eq:jointE}
% \mathbf{P}(\evec) &= \mathbf{P}(\etotx) \times \mathbf{P}(\etotz) \notag \\
% & = \frac{b_1^{a_1}}{\Gamma(a_1)}\etotx^{a_1-1}e^{-b_1\etotx} \times \frac{b_2^{a_2}}{\Gamma(a_2)}\etotz^{a_2-1}e^{-b_2 \etotz},
% \end{align}
% where $a_{i}$ and $b_{i}$ are parameters for each enzyme. 

To compute the mixture components with the LNA, we use the rate equations for the reactions in \eqref{eq:epi_PC_1}--\eqref{eq:epi_PE_2}; the full set of ODEs is listed in Eq. \eqref{eq:ODE2} in the Methods. As in the first example, by employing the conservation laws in \eqref{eq:cons2} and assuming rapid equilibrium of the complexes $\cx$ and $\cz$, the deterministic model can be further simplified to a 2-dimensional ODE:
\begin{align}
\begin{split}
\frac{\text{d}\px}{\text{d} t}  &=  f(P_1,P_2)  - g(P_1,P_2) -  \kconsx \px ,\\
\frac{\text{d}\pz}{\text{d} t}  &= g(P_1,P_2) -  \kconsz \pz,
\end{split}\label{eq:ODEred2}
\end{align}
where for ease of notation we have omitted the dependency on $\etotx$ and $\etotz$. The nonlinear functions in \eqref{eq:ODEred2} are
\begin{align}
\begin{split}
& f(P_1,P_2) =  \etotx\frac{\kpa \substrate - \kpb \px}{ 1 + \theta \pz^h + \substrate\slash\kma + \px\slash\kmb },\\
& g(P_1,P_2) =  \etotz\frac{\kpc \px -\kpd \pz}{1 + \px\slash\kmc + \pz\slash\kmd}, 
\end{split}
\end{align}
where $\theta=\kseq\slash\kunseq$ is the product-enzyme binding constant and the remaining parameters are defined as $\kpa =  \kcx \kfxx\slash(\kbx + \kcx)$, $\kpb =  \kbx \krx\slash(\kbx + \kcx)$, $\kpc  = \kcz\kfz\slash(\kbz + \kcz)$, $\kpd = \kbz \krz\slash(\kbz + \kcz)$, $\kma = \kcx \slash \kpa$, $\kmb = \kbx \slash \kpb$, $\kmc = \kcz \slash \kpc$ and $\kmd = \kbz \slash \kpd$. 

As in the previous example, the ODEs in \eqref{eq:ODEred2} correspond to the full model \eqref{eq:ODE2} rewritten in terms of both metabolites assuming that the enzyme-substrate reactions reach  equilibrium in a faster timescale than catalysis. This reduced model can be readily employed to obtain approximations for the mixture components with the LNA. If we denote as $\pmean=\pmean(\etotx,\etotz)$ the steady state solution of \eqref{eq:ODEred2}, we can write the Lyapunov equation as $A \Sigma + \Sigma A^{T} + B B^{T} = 0$ with 
$A$ and $BB^{T}$ given by
\begin{align}
   A &= \begin{bmatrix}
    \dfrac{\text{d}}{\text{d} \px} \left( f  - g  \right)-  \kconsx & \dfrac{\text{d}}{\text{d} \pz} \left( f - g \right)\\
    \dfrac{\text{d}g}{\text{d} \px}  &  \dfrac{\text{d}g}{\text{d} \pz} -  \kconsz 
    \end{bmatrix},\\
    B B^{T} &= \begin{bmatrix}
    f  +  g +  \kconsx \px &  -g \\
    -g  &  g +  \kconsz \pz
    \end{bmatrix},
\end{align}
where $f(\cdot)$, $g(\cdot)$, and their derivatives are evaluated at the steady state solution $\pmean(\etotx,\etotz)$. The Gaussian components of the mixture model are then
\begin{align}\label{eq:jointGaussian}
&\mathbf{P}(\px,\pz | \etotx,\etotz) = \notag\\
&\frac{1}{2\pi |\Sigma(\etotx,\etotz)|}e^{-\frac{1}{2}(P-\pmean(\etotx,\etotz))^{T}\Sigma^{-1}(P-\pmean(\etotx,\etotz))},
\end{align}
where $P=(P_1,P_2)^{T}$ and $|\cdot|$ is the matrix determinant. After combining the joint distribution of enzymes and the components into Eq.~\eqref{eq:GMM2}, we get a Gaussian mixture model for the joint marginal distribution of both metabolites:
\begin{align}\label{eq:jointPDF}
    & \mathbf{P}(P_1,P_2) = \notag\\ & K\sum_{x,y=0}^{\infty}\frac{\Gamma(a_1+x)\Gamma(a_2+y)}{\Gamma(x+1)\Gamma(y+1)}\lp\frac{b_1}{1+b_1}\rp^{x}\lp\frac{b_2}{1+b_2}\rp^{y}\times \notag \\
    & \frac{1}{|\Sigma(x,y)|}e^{-\frac{1}{2}(P-\pmean(x,y))^{T}\Sigma(x,y)^{-1}(P-\pmean(x,y))},
\end{align}
where $\pmean(x,y)$ and $\Sigma(x,y)$ need to computed numerically for each pair $(x,y)=(\etotx,\etotz)$ in the summation. The burst frequencies $a_i = k_{\text{tx},i}\slash\kdil$ and burst sizes $b_i =k_{\text{tl},i}\slash k_{\text{deg},i}$ are specific to each enzyme, and the normalisation constant is given by
\begin{align}
    K = \frac{1}{2\pi\Gamma(a_1)\Gamma(a_2)(1+b_1)^{a_1}(1+b_2)^{a_2}}.
\end{align}

\begin{figure}
\includegraphics{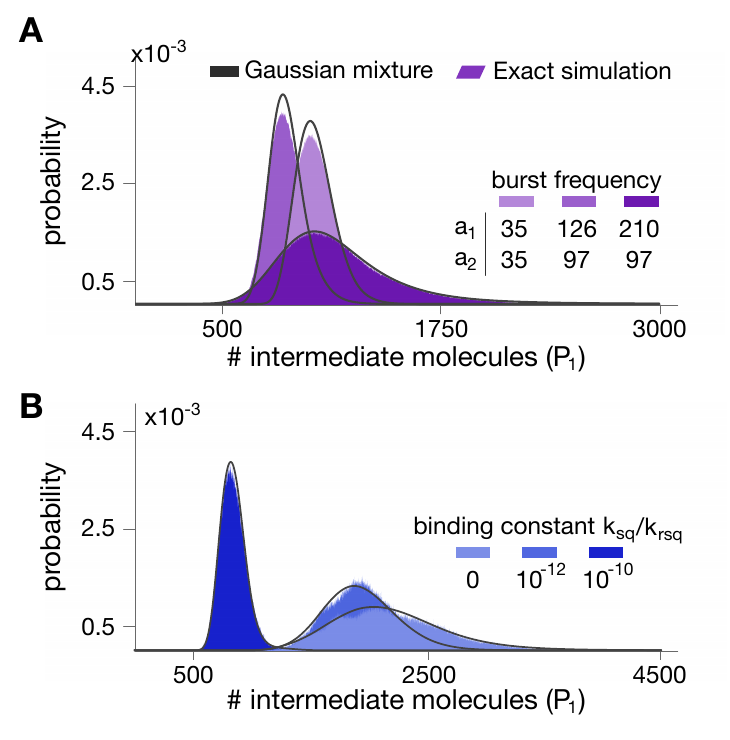}
\caption{\textbf{Stationary distributions for the intermediate metabolite in a two-step pathway with end-product inhibition.} The panels show the distribution of intermediate metabolite $\px$ for different combinations of parameter values. \textbf{(A)} Impact of enzyme bursting frequency $a_{1}$ and $a_2$. \textbf{(B)} Impact of binding constant between the first enzyme and the end-product. All parameter values can be found in Table \ref{tab:casestudy2} in the Methods.}
\label{fig:casestudy2a}
\end{figure}

To test the quality of the approximation, we numerically computed the mixture model in \eqref{eq:jointPDF} for various combinations of parameter values, shown in  \figref{fig:casestudy2a}. We observe that the mixture model offers an excellent approximation as compared to exact Gillespie simulations of the full model \eqref{eq:epi_PC_1}--\eqref{eq:epi_DIL}. We note that in this case, the full stochastic model has seven species and three different timescales, and therefore the runtime of Gillespie simulations are extremely long, in the order of several hours per run.

\begin{figure*}
\includegraphics{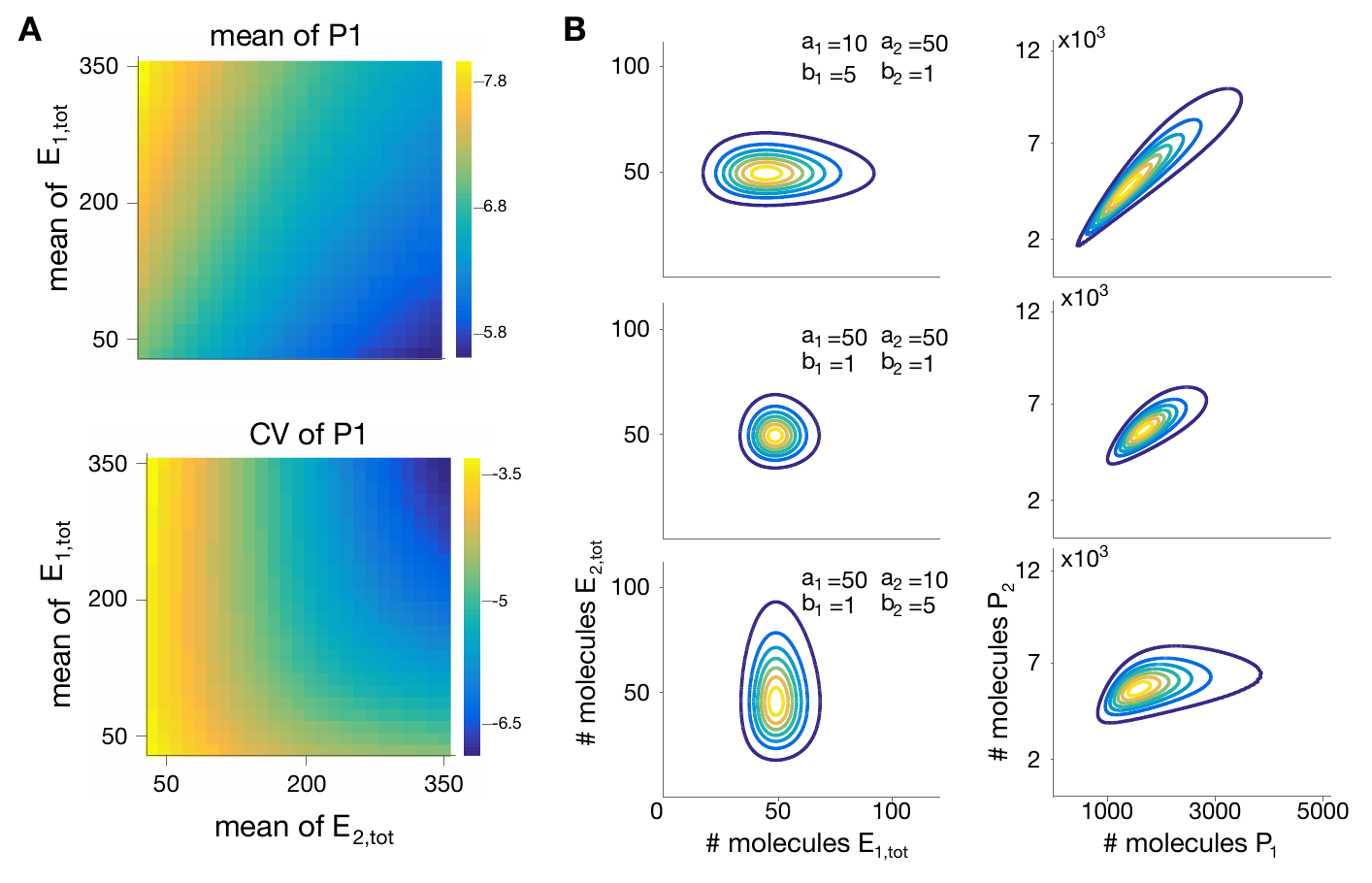}
\caption{\textbf{Impact of enzyme expression on metabolite distributions.} \textbf{(A)} For a wide sweep of the total enzyme expression levels, we observe changes in the mean abundances and coefficient of variation (CV) of the intermediate metabolite $\px$. Shown are the logarithmic values of mean and coefficient of variation. \textbf{(B)} Enzyme bursting parameters can strongly shape the metabolite distribution. All parameter values can be found in Table \ref{tab:casestudy2b} in the Methods.}
\label{fig:casestudy2b}
\end{figure*}

To further illustrate the utility of our method, we employed the mixture model to study the impact of parameter perturbations on the metabolite distributions. Without an analytical solution, such a study would require the computation of long Gillespie simulations for each combination of parameter values, which quickly become infeasible due to the long simulation time. In contrast, the mixture model provides a systematic way to rapidly evaluate the influence of model parameters on metabolite distributions. In \figref{fig:casestudy2b}A we show summary statistics of the marginal $\mathbf{P}(P_1)$ for various combinations of average enzyme expression levels. The results suggest that expression levels can have a strong impact on the mean and coefficient of variation of the intermediate metabolite. Moreover, in \figref{fig:casestudy2b}B we plot the distribution $\mathbf{P}(P_1,P_2)$ for combinations of bursting parameters. The results show that uncorrelated enzyme fluctuations can still result in correlated and skewed metabolite distributions.

%%%%%%%%%%%%%%%%%%%%%%%%%%%%%%%%%%%%%%%%%%%%%

\section{Discussion}
%rocked foundations of this paradigm,

Cellular metabolism has traditionally been assumed to follow deterministic dynamics. This paradigm results largely from the observation that cellular metabolites are highly abundant. However, recent data shows that single-cell metabolite distributions can display  substantial heterogeneity in their abundance across single cells\cite{Esaki2015, Bennett2009, Xiao2016, Mannan2017, Lemke2011, Yaginuma2014, Imamura2009, Paige2012, Ibanez2013}. It has also been shown that expression of metabolic genes is as variable as any other component of the proteome\cite{Taniguchi2010}, and thus in principle it is plausible that such enzyme fluctuations propagate to metabolites. These observations have begun to challenge the paradigm of metabolism being a deterministic process, suggesting that metabolite fluctuations may play a role in non-genetic heterogeneity. 

Here we described a new computational tool to predict the statistics of metabolite fluctuations in conjunction with gene expression. The method is based on a timescale separation argument and leads to a Gaussian mixture model for the stationary distribution of cellular metabolites. Computing distributions from this approximate model is substantially faster than through stochastic simulations, as these can be extremely slow due to the multiple timescales of metabolic pathways. Our technique can therefore be employed to efficiently explore the parameter space and predict the shape of metabolite distributions in different conditions. In earlier work we showed that the product of a single metabolic reaction can be accurately described by a Poisson mixture model\cite{Tonn2019}. Such approximation allowed the discovery of previously unknown regimes for metabolite distributions, including heavily tailed distributions and various types of bimodality and multimodality. The Poisson approximation, however, is bespoke to single reactions and not valid for more complex systems. In contrast, the Gaussian mixture model discussed here is more general and can  be applied to multiple kinetic mechanisms, more complex stoichiometries, as well as post-translational regulation. 

Another advantage of our approach is that the mixture weights can be computed offline from stochastic models for gene expression or single-cell expression data. The model is flexible in that it can readily accommodate gene expression models of various complexity. For the sake of illustration, in our examples we used the simple two-stage model for gene expression, but other models including gene regulation can also be employed\cite{Dattani2017}. Particularly relevant models are those that account for enzyme co-regulation, a widespread feature of bacterial operons\cite{Chubukov2014}, which translates into correlations between expression of different pathway enzymes and the resulting metabolite abundances.

In principle, most metabolic reactions satisfy the timescale separation as a result of their kinetics being much faster than the rate at which cells can synthesise new enzymes. 
However, throughout our examples we assumed that the metabolic substrate $S$, which is typically a carbon source or other precursor, remains constant. This case represents an abundant nutrient source with little fluctuations, but it is not adequate when substrates are lowly abundant or subject to stochastic fluctuations dictated by the environment. For example, depending on the timescale of such environmental fluctuations, the substrate can become another source of variability apart from enzyme expression\cite{Dattani2017}. In such cases, the timescale separation argument may not hold anymore and our theory needs to be revised to account for substrate fluctuations.

A number of works have sought to find links between fluctuations across layers of cellular organisation, such as gene expression, metabolism and cell growth\cite{Kiviet2014,Kotte2014,vanHeerden2014,Nikolic2017,Thomas2018}. But since measurement of metabolites in single cells remains technically challenging, there is pressing need for computational methods to predict fluctuations in cellular metabolites. Our proposed method provides a systematic approach for such task, paving the way for the generation of hypotheses on the molecular sources of metabolic heterogeneity.

\section{Methods}
\subsection{Model simulation}

Stochastic simulations were computed with Gillespie’s algorithm over long simulation times (several hours) corresponding to thousands of cell cycles. The ODE models and Lyapunov equations were solved in Matlab. In all examples, the negative binomial distribution for gene expression in \eqref{eq:negbin} was computed with its continuum approximation (Gamma distribution).

\subsection{Deterministic rate equations}
\paragraph{Reversible Michaelis Menten.} The full set of rate equations for the reversible reaction in \eqref{eq:revMM1}--\eqref{eq:revMM2} is:
\begin{align}
\begin{split}
    \frac{\text{d}\product}{\text{d} t} &=  \kcat \complexb - \krev \enzymeb \product - \kexp \product\\
    \frac{\text{d}\enzymeb}{\text{d} t} &=  -\kfor \substrate \enzymeb + \kback \complexb + \kcat \complexb - \krev \enzymeb \product,\\
    \frac{\text{d}\complexb}{\text{d} t} & = \kfor \substrate \enzymeb - \kback \complexb - \kcat \complexb + \krev \enzymeb \product. 
\end{split}\label{eq:ODE1}
\end{align}
To further reduce the above system of ODEs to Eq. \eqref{eq:ODEred1} in the main text, we can substitute the conservation relation in Eq. \eqref{eq:cons1}, \ie $\complex = \etotal -\enzyme$, and use the fact that the substrate-enzyme complex ($C$) typically equilibrates much faster than the product $P$, which means that $\text{d}C\slash\text{d}t\approx 0$ in the timescale of catalysis.

\paragraph{End-product inhibition.} The full set of rate equations for the reactions in \eqref{eq:epi_PC_1}--\eqref{eq:epi_PE_2} is:
\begin{align}
\begin{split}
\frac{\text{d}\px}{\text{d} t}  &= \kcx \cxb - \krx \exb \px  - \kfz \ezb \px + \kbz \czb - \kconsx \px\\
\frac{\text{d}\pz}{\text{d} t}  &= \kcz \czb - \krz \ezb \pz  -\kseq \exb \pz^h + \kunseq \seqb - \kconsz \pz.\\
\frac{\text{d}\exb}{\text{d} t}  &= -\kfx \substrate \exb + \left(\kbx + \kcx\right) \cxb - \krx \px \exb \\
& - \kseq \pz^h \exb + \kunseq \seqb,\\
\frac{\text{d} \cxb}{\text{d} t} &= \kfx \substrate \exb - \left(\kbx + \kcx\right) \cxb + \krx \px \exb, \\
\frac{\text{d} \seqb}{\text{d} t}  &= \kseq \pz^h \exb - \kunseq \seqb,\\
\frac{\text{d} \ezb}{\text{d} t}  &= -\kfz \px \ezb + \left(\kbz + \kcz \right) \cxb - \krz \pz \ezb,\\
\frac{\text{d} \czb}{\text{d} t} &= \kfz \px \ezb - \left(\kbz + \kcz \right) \cxb + \krz \pz \ezb
\end{split}\label{eq:ODE2}
\end{align}

As in the previous example, we can use the rapid equilibrium assumption and the conservation relations in \eqref{eq:cons2}, \ie $\etotx = \ex + \seq + \cx$ and $\etotz = \ez + \cz$, to simplify the 7-dimensional ODE in \eqref{eq:ODE2} to the 2-dimensional system in \eqref{eq:ODEred2} of the main text. 

\section*{References}
\bibliographystyle{unsrt}

 \newpage
 
 \begin{table*}
    \centering
    \begin{tabular}{c|cc|c}
    \multicolumn{4}{c}{\cellcolor{gray!20} \figref{fig:casestudy1}}\\
        $\delta$ & $0.00025 \text{ s}^{-1}$ & $\kback$ & $1000 \text{ s}^{-1}$\\
        $a$ & $\{ 25, 50, 120 \}$ & $\kcat$ & $3.6 \text{ s}^{-1}$\\
        $b$ & $1$ & $\krev$ & $0.01  \text{ s}^{-1}$\\
        $S$ & $3000 \text{ molecules}$ & $\kexp$ & $0.02 \text{ s}^{-1}$\\
        $\kfor$ & $1 \times S \text{ s}^{-1}$ & 
    \end{tabular}
    \caption{Parameter values for simulations in \figref{fig:casestudy1}.}
    \label{tab:casestudy1}
\end{table*}

\begin{table*}
    \centering
    \begin{tabular}{c|cc|c}
    \multicolumn{4}{c}{\cellcolor{gray!20}\figref{fig:casestudy2a}}\\
        $\delta$ & $0.00025 \text{ s}^{-1}$ & $\krx$ & $0.0001  \text{ s}^{-1}$\\
        $\kdegrx$ & $0.2 \text{ s}^{-1}$ & $\kconsx$ & $0.00025 \text{ s}^{-1}$\\
        $\kdegrz$ & $0.2 \text{ s}^{-1}$ & $\kfz$ & $ 1.5  \text{ s}^{-1}$\\
        $S$ & $3000 \text{ molecules}$ & $\kbz$ & $ 15000\text{ s}^{-1}$\\
        $\kfx$ & $20 \times S \text{ s}^{-1}$ & $\kcz$ & $ 150\text{ s}^{-1}$\\
        $\kbx$ & $15000 \text{ s}^{-1}$ & $\krz$ & $ 0.001  \text{ s}^{-1}$\\ 
        $\kcx$ & $22.5 \text{ s}^{-1}$ & $\kconsz$ & $ 0.15\text{ s}^{-1}$ \\
    \end{tabular}
        \begin{tabular}{c|c}
        \multicolumn{2}{c}{\cellcolor{gray!20} \figref{fig:casestudy2a}A}\\
        $a_1$ & $\{ 35, 126, 210 \}$ \\
        $a_2$ & $\{ 35, 97, 97\}$ \\
        $b_1$ & $1$\\
        $b_2$ & $1$\\
        $\kseq$ & $ 10^{-10} \text{ s}^{-1}$ \\
        $\kunseq$ & $ 1  \text{ s}^{-1}$ \\
        $h$ & $ 3$\\
    \end{tabular}
    \begin{tabular}{c|c}
    \multicolumn{2}{c}{\cellcolor{gray!20}\figref{fig:casestudy2a}B}\\
        $a_1$ & $80$\\
        $a_2$ & $80$\\
        $b_1$ & $1$\\
        $b_2$ & $1$\\
        $\kseq$ & $\{ 0, 10^{-10}, 10^{-12} \} \text{ s}^{-1}$ \\
        $\kunseq$ & $ 1  \text{ s}^{-1}$ \\
        $h$ & $ 3$\\
    \end{tabular}
    \caption{Parameter values for simulations in \figref{fig:casestudy2a}.}
    \label{tab:casestudy2}
\end{table*}

\begin{table*}
    \centering
    \begin{tabular}{c|c}
    \multicolumn{2}{c}{\cellcolor{gray!20}\figref{fig:casestudy2b}A}\\
        $a_1$ & $[10,100]$\\
        $a_2$ & $[10,100]$\\
        $b_1$ & $1$\\
        $b_2$ & $1$\\
        $\kseq$ & $ 10^{-10} \text{ s}^{-1}$ \\
        $\kunseq$ & $ 1  \text{ s}^{-1}$ \\
        $h$ & $ 3$\\ 
    \end{tabular}
    \begin{tabular}{c|c}
    \multicolumn{2}{c}{\cellcolor{gray!20}\figref{fig:casestudy2b}B}\\
        $a_1$ & $\{ 10, 50, 50 \}$\\
        $a_2$ & $\{ 50, 50, 10 \}$\\
        $b_1$ & $\{ 5, 1, 1 \}$\\
        $b_2$ & $\{ 1, 1, 5 \}$\\
        $\kseq$ & $ 0 \text{ s}^{-1}$ \\
        $\kunseq$ & $ 1  \text{ s}^{-1}$ \\
        $h$ & $ 3$\\
    \end{tabular}
    \caption{Parameter values for simulations in \figref{fig:casestudy2b}.}
    \label{tab:casestudy2b}
\end{table*}
% \onecolumngrid
% \appendix
% \input{appendix}

\end{document}